\documentclass[12pt]{article}

\usepackage[dvips]{graphicx,color}

\begin{document}

\begin{center}

{\Large {\bf Electron neutrino mass scale in spectrum \\

\vspace{0,3cm}

of Dirac equation with the 5-form flux term  \\

\vspace{0,3cm} 

on the AdS(5)xS(5) background}}

\vspace{2cm}

{Boris L. Altshuler}\footnote[1]{E-mail adresses: baltshuler@yandex.ru \& altshul@lpi.ru}

\vspace{0,5cm}

{\it Theoretical Physics Department, P.N. Lebedev Physical
Institute, \\  53 Leninsky Prospect, Moscow, 119991, Russia}

\vspace{2cm}

{\bf Abstract}

\end{center}

Dimensional reduction from 10 to 5 dimensions of the IIB supergravity Dirac equation written down on the AdS(5)xS(5) (+ self-dual 5-form) background provides the unambiguous values of bulk masses of Fermions in the effective 5D Randall Sundrum  theory. The use of "untwisted" and "twisted" (hep-th/0012378) boundary conditions at the UV and IR ends of the warped space-time results in two towers of spectrum of Dirac equation: the ordinary one which is linear in spectral number and the "twisted" one exponentially decreasing with growth of spectral number. Taking into account of the Fermion-5-form interaction (hep-th/9811106) gives the electron neutrino mass scale in the "twisted" spectrum of Dirac equation. Profiles in extra space of the eigenfunctions of left and right "neutrinos" drastically differ which may result in the extremely small coupling of light right neutrino with ordinary matter thus joining it to plethora of candidates for Dark Matter.

\vspace{0,5cm}

\newpage
\section{Presentation of Dirac equation including the Pauli-type term}

\qquad To receive the observed spectra of Fermions from the higher-dimensional theories is a long-standing problem. Introduction of Higgs scalar is a conventional Standard Model approach to generate masses of Fermi fields. However mass-like terms in Dirac equations in higher-dimensional theories may appear also because of interaction of Fermion with gauge fields (see e.g. Review \cite{Quiros}) or with $n$-form fields (\cite{Trip} and references therein). Thus the interesting task is to study the influence of the extra-dimensional Pauli type terms in the bulk Fermi field Lagrangian on the properties of mass spectra of Fermi excitations on different supergravity backgrounds. 

In the present paper we explore spectrum of D10 Dirac equation with the flux-generated bulk "mass term" in the Type IIB supergravity \cite{Arutyun}

\begin{equation}
\label{1}
\left(\Gamma^{M}D_{M}- \frac{i}{2 \cdot 5!} \Gamma^{M_{1}\ldots M_{5}} F_{M_{1}\ldots M_{5}}\right) \hat \lambda = 0,
\end{equation}
on the $AdS_{5}\times S^{5}$ (+ self-dual 5-form) background:

\begin{equation}
\label{2}
ds_{10}^{2}=e^{-2z/L}\eta_{\mu\nu}dx^{\mu}dx^{\nu} + dz^{2} + L^{2} d\Omega_{5}^{2},
\end{equation}

\begin{equation}
\label{3}
F_{0123z}=e^{-4z/L} \bar{Q}/L, \qquad F_{56789}=L^{4} \bar{Q},  \qquad   \bar{Q}=1.
\end{equation}

The value of the 5-form charge $\bar{Q}=1$ follows from the Einstein equations in 10 dimensions for the choice of normalization of the 5-form taken in the Type IIB supergravity action in \cite{Arutyun}:

\begin{equation}
\label{4}
S=\frac{1}{2k^{2}}\int\,d^{10}x\,\sqrt{-g}\left(R-\frac{4}{5!}F_{M_{1}\ldots M_{5}}F^{M_{1}\ldots M_{5}} + \dots \right),
\end{equation}
$k$ is gravity constant in 10 dimensions: $k=l_{s}^{4}$, $l_{s}$ is fundamental string length. We follow here the notations of \cite{Arutyun}: $M, N = 0,1\ldots 9$, $x^{M}=(x^{a}, y^{\alpha})$, $x^{a}=(x^{\mu},z)$ ($\mu=0,1,2,3$) and $y^{\alpha}$ are five angles of $S^{5}$ ($\alpha = 5,6,7,8,9$); hatted symbols, $\hat M$ etc., below are the corresponding flat indices. $\eta_{\mu\nu}$ in (\ref{2}) is metric of Minkowski space-time with signature {($-,+,+,+$)}.

D10 space-time is as ordinary orbifolded at the $UV$ and $IR$ boundaries given by the corresponding values of proper coordinate $z$:

\begin{equation}
\label{5}
z_{UV}=0 < z < \pi R = z_{IR},
\end{equation}
$AdS_{5}\times S^{5}$ space-time consists of two pasted copies with $Z_{2}$ symmetry imposed at its UV and IR ends. In the paper only bulk equations are explored, it is supposed that there are no additional surface terms of the Action which may influence the dynamics of Fermions.

The low-energy effective action (\ref{4}) makes sense if scale of curvature of space-time (\ref{2}) is essentially below the fundamental scale, i.e. if {$L\gg l_{s}$}. Standard dimensional reduction of Einstein term in (\ref{4}) with use of background metric (\ref{2}) gives the following expression for Planck Mass in 4 dimensions through length parameters $L$, $l_{s}$ (cf. e.g. \cite{Wolfe}):

\begin{equation}
\label{6}
M_{Pl}=\sqrt{\frac{\pi^{3}}{2}}\cdot \left(\frac{L}{l_{s}}\right)^{4}\frac{1}{L},
\end{equation}
here the exponentially small contribution from $z_{IR}$ limit of integration over $z$ in (\ref{4}) is omitted and value of volume of unit 5-sphere $\Omega_{5}=\pi^{3}$ is used.

$\hat \lambda$ in (\ref{1}) is a 32-component spinor, $D_{M}=i\partial_{M}{+} (1/4) \omega_{M}^{\hat A\hat B}\Gamma_{\hat A}\Gamma_{\hat B}$ is derivative including spin-connection, and the often used \cite{Mets}, \cite{Arutyun} representation for $32\otimes 32$ gamma-matrices $\Gamma ^{\hat M}$ is supposed:

\begin{eqnarray}
\label{7}
&&\Gamma^{\hat a}=\gamma^{a}\otimes \sigma^{1} \otimes I_{4}, \qquad \gamma^{a}=(\gamma^{\mu}, \, \gamma_{5}); \nonumber
\\
\\
&&\Gamma^{\hat\alpha}=-I_{4}\otimes \sigma^{2}\otimes \tau^{\alpha}, \qquad \tau^{\alpha}=(\tau^{i}, \, \tau_{5}), \nonumber
\end{eqnarray}
here $\gamma^{\mu}$, $\tau^{i}$ ($i=1,2,3,4$) are ordinary gamma-matrices in flat 4D Minkowski and Euclidian spaces correspondingly; $\gamma_{5}=-i\gamma^{0}\gamma^{1}\gamma^{2}\gamma^{3}$, $\tau_{5}=\tau^{1}\tau^{2}\tau^{3}\tau^{4}$. Chiral operator in 10 dimensions $\Gamma_{11}=\prod\nolimits_{0}^{9}\Gamma^{\hat M}= I_{4}\otimes \sigma^{3}\otimes I_{4}$; $\sigma^{1,2,3}$ are Pauli matrices.

As it was shown in \cite{Arutyun} flux term in (\ref{1}) projects out for one of chiral components of $\hat \lambda$ which is easily seen by direct calculation from (\ref{3}), (\ref{7}) (here world gamma-matrices $\Gamma^{\mu}= e^{z/L}\Gamma^{\hat \mu}$, $\Gamma^{z}=\Gamma^{\hat z}$, $\Gamma^{\alpha}=L^{-1}\Gamma^{\hat \alpha}$ are used):

\begin{equation}
\label{8}
\frac{i}{2\cdot 5!}\Gamma^{M_{1}\ldots M_{5}}F_{M_{1}\ldots M_{5}}= -\frac{\bar{Q}}{L}\sigma^{+}\otimes I_{16},
\end{equation}
$\sigma^{+}=(\sigma^{1}+i\sigma^{2})/2$ is a projection $2\otimes 2$ operator. Thus writing down Dirac equation (\ref{1}) for every of two 16-component chiral spinors $\lambda_{Q}$ we must take $Q=\bar{Q}=1$ for the right-handed chiral spinor \cite{Arutyun} and $Q=0$ for the left-handed one.

Again following \cite{Arutyun} we expand chiral spinor in a set of spherical harmonics of $S^{5}$:

\begin{eqnarray}
\label{9}
&& \lambda_{Q}(x^{\mu},z, y^{\alpha})= \sum_{n\ge 0}[\lambda_{Qn}^{+}(x^{\mu}, z)\otimes \chi_{n}^{+}(y^{\alpha})+ \lambda_{Qn}^{-}(x^{\mu}, z)\otimes \chi_{n}^{-}(y^{\alpha})],  \nonumber   
\\
\\
&& \qquad \tau^{\alpha}D_{\alpha}\chi_{n}^{\pm}(y^{\alpha})= \mp \frac{i}{L}\left(n+\frac{5}{2}\right)\chi_{n}^{\pm}(y^{\alpha}). \nonumber
\end{eqnarray}

Summing up these preliminaries the following Dirac equations in 5 dimensions $(x^{\mu},\, z)$ are received from (\ref{1}) for the 4-component spinors $\lambda_{Qn}^{\pm}$:

\begin{eqnarray}
\label{10}
&& \left[ e^{z/L}\gamma^{\mu}\partial_{\mu}+ \gamma_{5}\partial_{z}-\frac{2}{L}\gamma_{5} +  \frac{1}{L}\left(n+\frac{5}{2}+ Q\right) \right]\lambda_{Qn}^{+}(x^{\mu},z)=0, \nonumber
\\
\\
&& \left[ e^{z/L}\gamma^{\mu}\partial_{\mu}+ \gamma_{5}\partial_{z}-\frac{2}{L}\gamma_{5} -  \frac{1}{L}\left(n+\frac{5}{2}-Q \right) \right]\lambda_{Qn}^{-}(x^{\mu},z)=0,  \nonumber
\end{eqnarray}
$n=0,1,2\ldots$, $Q=1,\,0$ (we remind that $Q=1$ for the right-handed 16-component chiral spinor and $Q=0$ for the left-handed one). Term $(2/L) \gamma_{5}$ appears in (\ref{10}) from the spin-connection in $D_{\mu}$ in (\ref{1}) and reflects $z$-dependence of the warp factor in metric (\ref{2}). 

We'll see that most interesting (because of lesser value of coefficient in round brackets) is the case of second equation in (\ref{10}); thus in what follows Dirac equation for 4-component spinors $\lambda_{Qn}^{-}$ will be considered. Further separation of variables:

\begin{equation}
\label{11}
\lambda_{Qn}^{-}(x^{\mu},z)=(\lambda_{QnL}^{-}, \lambda_{QnR}^{-})= (\psi_{L}(x^{\mu})\, f_{L}(z), \,\psi_{R}(x^{\mu})\, f_{R}(z)),
\end{equation}
(indices $Q$, $n$, $-$ are omitted in the RHS of (\ref{11}); $\psi_{L}$, $\psi_{R}$ are the left and right components of Dirac spinor $\psi (x^{\mu})= (\psi_{L},\, \psi_{R})$ of mass $m$ governed by the ordinary Dirac equation in 4 dimensions $(\gamma^{\mu}\partial_{\mu}-m) \psi =0$) reduces equation (\ref{10}) for $\lambda^{-}_{Qn}$ to the following system for profiles $f_{L,R}(z)$ (for every $Q$, $n$):

\begin{eqnarray}
\label{12}
&& \left[\frac{d}{dz} - \frac{2}{L} - \frac{1}{L}\left(\nu+1/2\right)\right] f_{L} + m \, e^{z/L}f_{R}=0,  \nonumber   
\\
\\
&& \left[\frac{d}{dz} - \frac{2}{L} + \frac{1}{L}\left(\nu+1/2\right)\right] f_{R} - m \, e^{z/L}f_{L}=0, \nonumber
\end{eqnarray}
parameter $\nu = n+2-Q$ essentially determines the looked for spectra of $m$; for $Q=1$ (i.e. for Fermions which "feel" the flux) $\nu = 1,{}2\ldots $, and in case $Q=0$ we have $\nu = 2,{}3\ldots$

Equations (\ref{12}) are typical in the Randall-Sundrum type models when bulk Dirac mass term is included in the Fermi field Lagrangian \cite{Neubert}, \cite{Gher1}, \cite{Gher2}, \cite{Gher3}. However, contrary to these papers were value of bulk Dirac mass which determines the physically important parameter $\nu$ in (\ref{12}) was taken "by hand", here we rely upon well grounded supergravity approach which gives definite values of $\nu$.

\section{Boundary conditions, two towers of spectra of Fermi excitations and "seesaw" scale without seesaw mechanism}

Solution of system (\ref{12}) is a linear combination of Bessel and Neuman functions \cite{Gher1}-\cite{Gher3}:

\begin{eqnarray}
\label{13}
&& f_{L}(z)= e^{5z/2L}\left[AJ_{\nu}(\tau) + B N_{\nu}(\tau) \right],  \nonumber   
\\
\\
&& f_{R}(z)= e^{5z/2L}\left[AJ_{\nu+1}(\tau) + B N_{\nu+1}(\tau) \right],  \nonumber
\end{eqnarray}
where $\tau = mLe^{z/L}$; the slice of the $AdS_{5}\times S^{5}$ space-time is given by the interval $\tau$ (see (\ref{5})):

\begin{equation}
\label{14}
\tau_{UV}=mL < \tau < mL e^{\pi R} = \tau_{IR},
\end{equation}
$A,\,B$ in (\ref{12}) are constants determined from the boundary and normalization conditions. Boundary conditions give also the spectra of $m$.

At the reflection of coordinate $z$, $P_{z}$, D10 spinor $\hat \lambda$ transforms as (see e.g. \cite{Horava}):

\begin{equation}
\label{15}
P_{z}\hat \lambda (z)= \Gamma^{\hat z} \hat \lambda (-z).
\end{equation}

According to (\ref{7}) $\Gamma^{\hat z}=\gamma_{5}\otimes \sigma^{1}\otimes I_{4}$, hence $z$-reflection interchanges right-handed and left-handed chiral components of D10 spinor. It would be mistake however to think that this reflection interchanges chiral component interacting with flux (the right-handed one according to (\ref{8})) and the non-interacting one. The point is that electric and magnetic parts of the 5-form behave opposite under $z$-reflection: electric flux in (\ref{3}) is odd under reflection whereas magnetic one is even. Because of it reflection changes $\sigma^{+}$ to $-\sigma^{-}$ in (\ref{8}) ($\sigma^{-}=(\sigma^{1}-i\sigma^{2})/2$), and it is the left-handed 16-component chiral spinor which "feels" flux on the $Z_{2}$-symmetric pasted half of $AdS_{5}\times S^{5}$ space-time (\ref{2}).

Thus $Z_{2}$-symmetry adjustment at the orbifold points $z=z_{*}$ ($z_{*}=0, \pi R$) of the right-handed (left-handed) chiral component of D10 spinor $\hat \lambda$ "living" on one of pasted copies of slice of $AdS_{5}\times S^{5}$ space-time with the left-handed (right-handed) component on the other pasted copy gives the ordinary in the RS-type models boundary conditions for the left and right profiles (\ref{13}):

\begin{equation}
\label{16}
f_{R}(z_{*}) = \eta f_{R}(z_{*}), \qquad  f_{L}(z_{*})= - \eta f_{L}(z_{*}), \qquad \eta = \pm 1.
\end{equation}

We take for definiteness $\eta =1$ at the UV end and following \cite{Gher1}, \cite{Gher2}, \cite{Gher4} consider two types of boundary conditions: the usual "untwisted" one (when $\eta =1$ also at the IR end) and "twisted" one ($\eta =1$ at the UV end, $\eta = -1$ at the IR end). These conditions determine two essentially different towers of the eigenvalues of Dirac equation (\ref{1}). The "twisted" boundary condition corresponds to breaking supersymmetry by the Scherk-Schwarz mechanism \cite{SchSch}, \cite{Quiros} and its application for receiving small gravitino mass in the warped models was first proposed, as to our knowledge, in \cite{Gher2}. Thus let us consider:

\vspace{0,3cm}

{\it{"Untwisted" boundary conditions}}: 

$f_{R}(z_{*})=f_{R}(z_{*})$, $f_{L}(z_{*})=-f_{L}(z_{*})$ at both UV and IR boundaries. This means $f_{L}(0)=f_{L}(\pi R)=0$ which according to (\ref{13}) gives the spectral condition:

\begin{equation}
\label{17}
\frac{J_{\nu}(\tau_{UV}) }{N_{\nu}(\tau_{UV})}= \frac{J_{\nu}(\tau_{IR})}{N_{\nu}(\tau_{IR})},
\end{equation}
$\tau_{UV}$, $\tau_{IR}$ are given in (\ref{14}). For $mL \ll 1$, $R/L \gg 1$ (i.e. when $\tau_{UV} \ll 1$ and $\tau_{IR}$ is of order 1) solution of (\ref{17}) is given by simple formula \cite{Gher1}:

\begin{eqnarray}
\label{18}
&&m_{q,n}^{\it{untw}}\simeq \left(q+ \frac{\nu}{2}-\frac{3}{4}\right)\,\frac{\pi}{L}\, e^{-\pi R/L}=   \nonumber
\\ 
\\  
&&\left(q+\frac{n}{2}+\frac{1}{4}-\frac{Q}{2}\right)\,\sqrt{\frac{2}{\pi}}M_{Pl}\left(\frac{l_{s}}{L}\right)^{4}e^{-\pi R/L}\cong M_{EW}, \nonumber
\end{eqnarray}
where $q=1,2,3\ldots$, $n=0,1,2\ldots$, formula (\ref{6}) was used to express $L^{-1}$ through $M_{Pl}$.

Physically mass scale in the RHS of (\ref{18}) must be of order of the electro-weak scale $M_{EW}$; its relation to the Planck scale ("first mass hierarchy") is basically given by the small Randall-Sundrum exponent $e^{-\pi R/L}$ in (\ref{18}), although in the model under consideration it also depends on relation of fundamental string length $l_{s}$ to the scale $L$ of the Type IIB supergravity solution (\ref{2}). 

Since $\nu=n+2-Q >0$ the profiles of eigenfunctions (\ref{13}) of massive modes are essentially concentrated in vicinity of the IR end of the slice of $AdS_{5}\times S^{5}$ space-time. The "untwisted" boundary conditions permit also zero-mode ($m=0$) solutions of system (\ref{12}): $f_{L}\equiv 0$, $f_{R}\sim \exp [(-\nu +3/2)z/L]$.

\vspace{0,3cm}

{\it{"Twisted" boundary conditions}}: 

$f_{L}(z_{UV})=-f_{L}(z_{UV})$, $f_{R}(z_{IR})=-f_{R}(z_{IR})$, i.e. $f_{L}(0)=f_{R}(\pi R)=0$.
In this case spectral condition for solutions (\ref{13}) looks as \cite{Gher2}:

\begin{equation}
\label{19}
\frac{J_{\nu}(\tau_{UV}) }{N_{\nu}(\tau_{UV})}= \frac{J_{\nu+1}(\tau_{IR}) }{N_{\nu+1}(\tau_{IR})},
\end{equation}

There are no zero modes in this case. Spectral equation (\ref{19}) possesses the "first hierarchy" massive modes with eigenvalues of type (\ref{18}), but it also gives the "inverse tower" of extremely small values of $m_{Q,n}^{\it{tw}}$ exponentially decreasing with growth of spectral number $n$:

\begin{equation}
\label{20}
m_{Q,n}^{\it{tw}}=\frac{2\sqrt{n+2-Q}}{L}\, e^{-(n+3-Q)\pi R/L},
\end{equation}
which is received from (\ref{14}), (\ref{19}) with account that in this case both arguments in (\ref{19}) are much less then one ($\tau_{UV}=mL \ll 1$ and $\tau_{IR}=mLe^{\pi R/L} \ll 1$). We also inserted $\nu = n+2-Q$ in (\ref{20}).

The highest value of $m_{Q,n}^{\it{tw}}$ is achieved when Fermion interacts with flux, i.e. at $Q=1$, $n=0$ in (\ref{20}):

\begin{equation}
\label{21}
m_{1,0}^{\it{tw}}=\frac{2}{L}\, e^{-2\pi R/L}=\left(\frac{L}{l_{s}}\right)^{4}\frac{M_{EW}^{2}}{M_{Pl}}.
\end{equation}
In deriving the RHS of (\ref{21}) we expressed $L^{-1}$ and $e^{-\pi R/L}$ through $M_{Pl}$ and $M_{EW}$ from (\ref{6}), (\ref{18}) and omitted coefficient of order one. For the choice $M_{EW}=1TeV$, $(L/l_{s})^{4}= 10^{3}$ (\ref{21})  gives the mass scale of order of mass of electron neutrino. 

Theory surely must be more elaborated. The goal of the paper is to demonstrate interesting potential possibilities of the supergravity models, and to demonstrate the importance of presence of Pauli type terms in the bulk Dirac equations. In fact, let us calculate the next spectral value of tower (\ref{20}) which is received for $Q=0$, $n=0$ (or equivalently for $Q=1$, $n=1$):

\begin{equation}
\label{22}
m_{0,0}^{\it{tw}}=m_{1,1}^{\it{tw}}=\frac{\sqrt{8}}{L}\, e^{-3\pi R/L}=\left(\frac{L}{l_{s}}\right)^{8}\frac{M_{EW}^{3}}{M_{Pl}^{2}}.
\end{equation}
In the absence of the 5-form term in (\ref{1}) this would be the highest value of spectrum (\ref{20}) and physically promising "seesaw" combination $M_{EW}^{2}/M_{Pl}$ like in the RHS of (\ref{21}) would not appear in the spectrum of Dirac equation. It must be noted that mass scale $M_{EW}^{2}/M_{Pl}$ is received here without any reference to large right neutrino mass and standard seesaw mechanism (cf. \cite{Neubert}).

\section{Profiles of wave functions and light right neutrino as a candidate for Dark Matter}

Let us look at the profiles of eigenfunctions (\ref{13}) of "twisted" modes. With use of boundary conditions $f_{L}(0)= f_{R}(\pi R)=0$, inserting expression for $m$ given in (\ref{20}), and taking into account that in this case argument of cylinder functions in (\ref{13}) is small ($\tau \ll 1$) in all region (\ref{14}), it is easy to receive the simple approximate expressions for "twisted" eigenfunctions (\ref{13}):

\begin{eqnarray}
\label{23}
&& f_{L}^{\it{tw}}(z)= N_{\nu}\,e^{5z/2L}\sinh\left(\frac{\nu z}{L}\right),  \nonumber   
\\
\\
&& f_{R}^{\it{tw}}(z)= - N_{\nu}\sqrt{\nu}\,e^{5z/2L}\sinh\left[(\nu +1)\frac{(\pi R -z)}{L}\right], \nonumber
\end{eqnarray}
where $N_{\nu}$ is the normalization factor, $\nu = n+2-Q$.

From (\ref{23}) it is immediately seen that "twisted" profile $f_{L}(z)$ of the left component of 4-spinor $\psi (x^{\mu})$, see (\ref{11}), is concentrated near the IR end of the warped space-time (\ref{2}), whereas profile $f_{R}(z)$ of the right component is located near the UV end. This must result in essential difference in interactions of the left and right spinors with massive modes of other fields which profiles in extra space are concentrated near IR end of the higher-dimensional space-time.

In the extra-dimensional theories the strength of interaction of modes of different fields depends on overlapping of their wave functions in extra space. That is why universality of electric charge is achieved in these theories only if zero-mode of electro-magnetic field is constant in extra space. The same is true of course for the interaction of matter with gravitational field, the constancy of its zero-mode was supposed in deduction of expression (\ref{6}) for Planck mass in 4 dimensions. 

If Fermions considered above are neutral, then exponential suppression of the overlapping in extra space of wave function of right spinor with wave functions of the ordinary matter modes trapped on the IR brain or "living" in the bulk in vicinity of the IR end will make right neutral spinor unobservable even if its mass is small. Thus in this approach there is no need to suppose the extra large mass of right neutrino as an explanation of its unobservability in experiments; "twisted" right neutrino (\ref{11}) join the plethora of candidates for Dark Matter because of its extremely weak interaction with all "ordinary" matter but gravity.

\section{Discussion}

\qquad In \cite{Altsh08} it was shown that background flux "living" in extra sub-space makes massive Kaluza-Klein gauge field which corresponds to the isometrty group of this sub-space - in direct analogy with the standard Higgs mechanism. It is expected that background flux generating the Pauli type terms in the extra-dimensional Fermi fields' equations may also substitute Higgs scalar in forming the observed spectra of Fermions. The advantage of the supergravity models is in unambiguity of these terms in the Lagrangian of Fermi fields in high-dimensions.

The preliminary results of this paper surely may be generalized to the consideration of influence of fluxes on spectra of Fermions e.g. in the model of Klebanov-Strassler throat \cite{KS} in the Type IIB supergravity.

But perhaps the backgrounds of throat-like solutions in the Type IIA supergravity will prove to be even more promising for this direction of thought. In particular it would be interesting to explore if the naturally appearing in these models 2-form background flux quickly decreasing upward from the IR end of the throat (cf. \cite{Altsh08}, \cite{Altsh07}) may provide the intermediate mass scales of the Standard Model generations in spectra of neutral and charged Fermions calculated on this background. 

The ambitious goal is of course to receive the observed spectra of particles of Standard Model as a solution of classical equations in higher dimensions. The goal does not look too crazy in view of miraculous successes of dual-holography business in calculating spectra of bound states in QCD in its strong coupling limit from solutions of classical higher dimensional supergravity equations.

\qquad 

\section*{Acknowledgements} Author is grateful to R.R. Metsaev for valuable consultations. This work was supported by the Program for Supporting Leading Scientific Schools (Grant LSS-1615.2008.2).


\begin{thebibliography}{99}

\bibitem{Quiros}M. Quiros, {\it{"New Ideas in Symmetry Breaking"}}, Boulder 2002, Particle physics and cosmology, 549-601, hep-ph/0302189. 
\bibitem{Trip}P.K. Tripathy and S.P. Trivedi, {\it {"D3 Brane Action and Fermion Zero Modes in Presence of Background Flux"}}, JHEP 06 (2005) 066, hep-th/0503072.
\bibitem{Arutyun}G.E. Arutyunov. and S.A. Frolov, {\it{"Quadratic action for type IIB supergravity on $AdS_{5}\times S^{5}$"}}, JHEP 08 (1999) 024, hep-th/9811106.
\bibitem{Wolfe}O. De Wolfe O. and S.B. Giddings, {\it{"Scales and hierarchies in warped compactifications and brane worlds"}}, Phys. Rev. D {\bf 67}, 066008 (2003), hep-th/0208123.
\bibitem{Mets}R.R. Metsaev and A.A. Tseytlin, {\it{"Type IIB supergravity action in AdS(5) x S(5) background"}}, Nucl.Phys. {\bf B 533}, 109 (1998), hep-th/9805028.
\bibitem{Neubert}Y. Grossman and M. Neubert, {\it{"Neutrino masses and mixings in non-factorizable geometry"}}, Phys. Lett. {\bf{B474}} (2000) 361, hep-ph/9912408.
\bibitem{Gher1}T. Gherghetta and A. Pomarol, {\it{"Bulk fields and supersymmetry in a slice of AdS"}}, Nucl. Phys. {\bf B 586} (2000) 141, hep-th/0003129.
\bibitem{Gher2}T. Gherghetta and A. Pomarol, {\it{"A warped supersymmetric standard model"}}, Nucl. Phys. {\bf{B 602}}, 3 (2001), hep-th/0012378.
\bibitem{Gher3}T. Gherghetta, K. Kadota and M. Yamaguchi, {\it{"Warped Leptogenesis with Dirac Neutrino Masses"}}, hep-ph/0705.1749.
\bibitem{Horava}P. Horava and E. Witten, {\it{"Eleven-dimensional supergravity on a manifold with boundary"}}, Nucl.Phys. {\bf{B475}} (1996) 94-114, hep-th/9603142.
\bibitem{Gher4}T. Gherghetta and A. Pomarol, {\it{"A Stuckelberg formalism for the gravitino from warped extra dimensions "}}, hep-th/0203120.
\bibitem{SchSch}J. Scherk J.H. Schwarz, {\it{"Spontaneous breaking of supersymmetry through dimensional reduction"}}, Phys. Lett.{\bf{B82}} (1979) 60; {\it{"How to get masses from extra dimensions?"}}, Nucl. Phys. {\bf{B153}} (1979) 61.
\bibitem{Altsh08}B.L. Altshuler, {\it{Higgs mechanism from fluxes and two mass hierarchies in the "fat" throat solution of Type IIA supegravity}}, hep-th/0811.1486.
\bibitem{KS}I.R. Klebanov and M.J. Strassler, {\it{"Supergravity and a confining gauge theory: Duality cascades and $\chi SB$-resolution of naked singularities"}}, JHEP {\bf 0008}, 052 (2000), hep-th/0007191.
\bibitem{Altsh07}B.L. Altshuler, {\it{"Potential for the slow-roll inflation, mass scale hierarchy and dark energy from type IIA supergravity"}}, JCAP 09(2007) 012, hep-th/0706.3070.
\end{thebibliography}
\end{document}